\documentclass[12pt,a4paper]{article}

\usepackage{graphicx}
\usepackage{amssymb}
\usepackage{type1cm}
\usepackage{helvet}
\usepackage{courier}
\usepackage{makeidx}
\usepackage{url}
\usepackage{verbatim}
\usepackage{amsmath}
\usepackage{algorithm}
\usepackage{algorithmic}

\usepackage[margin=1in]{geometry}

\title{Scalable and Energy-Efficient Predictive Data Collection in Wireless Sensor Networks with Constructive Interference}

\author{
  Conor Muldoon \\
  Department of Computing and Mathematics, \\
  Chester Street, Manchester, M1 5GD, United Kingdom \\
  \texttt{c.muldoon@mmu.ac.uk}
}

\date{August 2025}

\begin{document}

\maketitle

\begin{abstract}
A new class of Wireless Sensor Network has emerged whereby multiple nodes transmit data simultaneously, exploiting constructive interference to enable data collection frameworks with low energy usage and latency. This paper presents STAIR (Spatio-Temporal Activation for Intelligent Relaying), a scalable, resilient framework for Wireless Sensor Networks that leverages constructive interference and operates effectively under stringent resource constraints. Using constructive interference requires all nodes to transmit the same packet at the same time, thus, only one source node can send data per time slot. STAIR uses coarse-grained topology information to flood a selected subset of the network, relaying sensor readings from individual nodes during their allocated time slots. A submodular optimisation algorithm with proven quality bounds determines near-optimal sensor activation locations and times, aiming to minimise the sum of mean squared prediction errors from a multiple multivariate linear regression model, which is used to estimate values at unselected locations and times. This framework has been extensively validated on a real-world testbed deployment.
\end{abstract}


\section{Introduction}

\label{intro}

Wireless Sensor Networks are increasingly being designed to support high-efficiency, low-latency communication enabled by new architectures and protocols that exploit advances in concurrent transmission techniques. This trend has grown rapidly in recent years \cite{baddeley2023understanding}. When multiple nodes transmit the same packet with a temporal displacement of less than 0.5 microseconds, there is high probability they will interfere constructively. Flooding-based architectures such as Glossy \cite{ferrari2011efficient,chang2018constructive} and Ripple \cite{yuan2015ripple} leverage this property to achieve highly efficient network-wide flooding and precise time synchronization. In many-to-one data collection scenarios, using flooding as a communication primitive requires all nodes to participate in every flood, activating to transmit data from a given source node. While this may ostensibly seem inefficient, frameworks like the Low Power wireless Bus (LPB) \cite{ferrari2012low,vachtsevanou2023embedding} demonstrate that it can be significantly more efficient than state-of-the-art contention-based protocols \cite{abba2024data}, including the Collection Tree Protocol \cite{gnawali2013ctp,gnawali2009collection,yuan2022collection}, and traditional TDMA-based methods \cite{sgora2015survey}. The STAIR framework described in this paper enhances the performance of Glossy and LPB for many-to-one data collection by selecting only subsets of nodes to participate in flooding. There are two aspects to subset selection: (1) a data collection method that uses constructive interference and subsets of nodes for relaying data and (2) a sensor placement and scheduling algorithm that incorporates both spatial and temporal network variance to determine near-optimal sensor activation times and locations. 

STAIR is intended for use in applications that operate over a spatial extent whereby various parts of the network are correlated to differing degrees. Given that sensors cannot be placed at every point within the space, statistical models should be adopted to determine the best locations and times to activate sensors such that the value of the information produced by the network and the predictive capabilities of the locations and times chosen are optimised. Sensor network applications that concern uncorrelated point information, such as detecting whether appliances are active in an Internet of Things scenario, shall not be considered here. 

When using constructive interference for flooding, it is necessary that all nodes in the network transmit the same packet. STAIR uses a greedy submodular optimisation algorithm with provable quality guarantees (see Section \ref{disc}) to determine which sensor location to transmit data from at a given time. Data from the selected sensor is used in conjunction with data from other sensors in a window, or series of time slots, to make predictions for locations where there are inactive sensors or for other locations of interest where there are no sensors placed. Predictions are made, using multiple multivariate linear regression, in each time slot of the window barring the sensors selected to be active in the time slots. The sensor placement and scheduling algorithm chooses the sensor location that minimises the sum of the mean squared prediction errors given the previously chosen sensor locations within the window to determine a sensor activation or transmission schedule. Within the schedule, sensors do not have a homogeneous sampling rate in that algorithm selects informative sensors to be activated more frequently. 

For many-to-one data collection, STAIR uses a subset of nodes to participate in flooding and coarse grained topology information to reduce total energy consumption whilst remaining robust to node failures and dynamics. Reducing the total energy consumption enables a much greater number of nodes to remain active for longer in networks that use flooding as a communication primitive. This is because if all nodes have the same workload, which is the case with flooding architectures, nodes closest to the sink will not necessarily deplete their energy resources first. The reason for this is that, in practice, the discharge rates of batteries vary. Thus, with approaches such as Glossy, nodes far from the sink will sometimes fail first in that they will have the same workload as nodes close to the sink along with variable rates of discharge. Traditionally, protocols were often not optimised for this type of problem as, with traditional protocols, nodes close to the sink had a much higher workload and would typically fail first. Thus, in contrast to the approach discussed in this paper, minimizing the total energy consumption of such a network would not make a substantial difference in relation to the amount of data received at the sink. 


To summarise, the main contributions of this paper are: a sensor placement and scheduling algorithm that provides improvements of up to 59\% over related research, a many-to-one communications framework that reduces the energy overhead and leads to an average improvement of over 67\% in terms of the number sensor readings received at the sink, and the evaluation of the algorithms on a real testbed and comparison with \cite{ferrari2011efficient,krause2008near,sakiyama2016,Clark2016} (see Section \ref{results}).

\section{Related Research}
\label{rr}

Prior research on sensor placement \cite{krause2008near,krause2008robust,muldoon2025sensor} has made use of submodular optimisation and selection criteria, such the mutual information and conditional entropy of Gaussian processes and the mean squared prediction error to achieve near-optimal performance in certain settings. Furthermore, submodular optimisation has been adopted in determining sensor placements for optimal Kalman filtering \cite{tzoumas2016sensor} and in maximising the Fisher information \cite{liu2023greedy}.
The algorithm discussed in Section \ref{algo} differs from prior research that uses greedy submodular optimisation in that a transmission scheduled is produced in conjunction to determining locations to activate sensors. It should be noted that sensor data are correlated both in space and in time. With STAIR, temporal and spatial considerations are taken into account, rather than just spatial, in determining the right sensor location at the right time to transmit data from. This is in contrast to prior research that determines the set of locations to activate sensors without regard to the frequency of sensor activation or temporal displacement or that makes the assumption that sensors should have a homogeneous sampling rate. The algorithm discussed in Section \ref{algo} jointly optimises over both sensor activation times and locations. Determining the locations and times to activate sensors in the transmission schedule is a form of subset selection problem, which, in general, is NP-hard. Often heuristics are used to address NP-hard problems, but with no guarantees with regard to the performance of such approaches. If the objective function, such as maximising the reduction in the mean squared prediction error, is submodular or approximately submodular (see Section \ref{disc}), however, which is the case in STAIR, using a greedy algorithm will perform close to optimal and with bounded quality guarantees. 

In terms of data transmission, STAIR operates in a similar manner to the Glossy flooding architecture \cite{ferrari2011efficient} and other more recent methods that make use of concurrent transmissions \cite{zimmerling2020synchronous,baddeley2023understanding}. With Glossy, no topology or routing information is used, rather, all data is flooded to all nodes in the network. This removes the need for a protocol to maintain or update a routing tree in that data is effectively transmitted through all paths and all nodes participate in all floods. With STAIR, however, only a subset of nodes within the network activate to transmit data from a given node. This reduces the total energy consumption of nodes within the network and increases the scalability of the architecture in many-to-one data collection scenarios.

One of the advantages of Glossy is that it does not require a routing tree or topology information for data transmission. With STAIR, topology information is used, but concurrent transmissions and multiple paths are also used to achieve similar results to Glossy, but with a lower energy consumption. The topology information is quite coarse grained, so it does not significantly degrade over time.  This enables STAIR, after an initial profiling phase, to continue operating over extended periods of time without the need to perform profiling again and with an improvement in the amount of data received.

Ripple \cite{yuan2015ripple} improves on the performance of Glossy by introducing pipeline transmissions on multiple channels and adding error coding, but does not consider flooding a subset of the network. It represents an orthogonal approach and could be used in conjunction with STAIR.

\section{Sensor Placement and Scheduling}
\label{select}

In deploying sensor networks, a decision must be made with regard to where to place sensors and with regard to which of the selected locations to transmit data from at a given time or in a given flood in frameworks that make use of flooding as a communication primitive. Sensors will often be correlated in different parts of the network and the predictive capabilities of the sensor locations chosen will vary. In this section, a placement and scheduling algorithm is discussed that determines the best locations and times for sensor activation with regard to the linear estimation of unselected locations and times or other locations of interest where there are no sensors placed. The algorithm chooses locations and times that minimise the approximate Mean Squared Prediction Errors (MSPEs) in creating a sensor activation or transmission schedule. This takes into account the trade-off between minimising the correlation between the current selection and previous selections and maximising the correlation between the current selection and the unselected set or locations and times of interest.

There is a limitation with prior research into sensor placement algorithms \cite{krause2008near,Clark2016,muldoon2025sensor,shamaiah2010greedy,sakiyama2016} in that it does not take into account the time at which sensing occurs when making decisions with regard to the locations that sensors should be activated. The following example illustrates the drawback of such an approach. Suppose there are 6 potential locations to activate sensors named A, B, C, D, E, F, but there are only 4 sensors available or owned by the application developer. A simple way of creating a sensor activation schedule would be to use a pre-existing sensor placement algorithm, such as \cite{krause2008near,sakiyama2016,Clark2016}, to choose a set of sensor locations and then to use round robin to schedule each sensor to activate in turn. With round robin, if locations A, B, D, and F were chosen by the placement algorithm, the schedule would be to activate A at time 1, B at time 2, D at time 3, F at time 4, A at time 5, and so forth. That is, once 4 time slots have elapsed, the sequence repeats.  Suppose A provided significantly better sensor readings and predictions than the other sensors for the area being monitored, however. In this scenario, it will be better to schedule A more frequently than the other sensors. 

With STAIR, sensor readings are used not only to provide information about the location at which they were sensed, but also to make predictions for other locations at specific times within a window or series of time slots. So, for instance, with a schedule of A, B, A, D, F, the sensor readings from location A at time 1, B at time 2, A at time 3, and so forth, would be used to make predictions for the entire set of locations A, B, C, D, E, F for each time slot within the window\footnote{Predictions are not made for the locations and times at which the sensors were activated, as this information is known. For instance, in this example, predictions would be made for B for times 1, 3, 4, 5 but not for time 2.}. This differs from prior work in that at each time slot (1) only a single sensor is selected to transmit data from, which is necessary when using constructive interference for transmission, and (2) sensors are still available for selection regardless of whether they have been selected to activate at prior time slots within the window. As such, rather than choosing a fixed subset of sensors to activate, the algorithm determines the time and frequency of sensor activation, which represents a transmission schedule. 

One objection to this might be that, in sensor networks, data from a number of sensors will often be transmitted at the same time, but this is not the case in situations whereby efficient flooding communication primitives are adopted. 

\subsection{Profiling and Prediction}
\label{propre}
Prior to the execution of the sensor placement and scheduling algorithm, there is an initial profiling phase whereby sensor values are received for all nodes in the network. This data is then used by the algorithm to construct multiple multivariate linear regression models for prediction. Given that there can only be one initiator or source node per flood or time slot, the goal is to determine which location at which times represent the best choice with regard to the estimator. To minimise the MSPEs, the placement and scheduling algorithm optimises over the sample mean of the squared difference of the observed values of a subset of the training data and the predicted values of the subset, which are predicted using a least squares estimator that was fitted to the training data. Once the algorithm completes execution, the least squares estimator, associated with the selected set of sensors, is used to make predictions on the test data or data coming from the network. For the predicted values of a least squares estimator $\hat{o}$ and the observed values $o$ of the training data, the MSPE is approximated as follows: $
\operatorname{MSPE}\approx\frac{1}{n}\sum_{i=1}^n(\hat{o_i} - o_i)^2
\label{mspe}
$

\begin{algorithm}
\caption{Sensor Placement and Scheduling}
\label{cde2}
\begin{algorithmic}[1]
\STATE $t \gets$ number time slots in a window
\STATE $ k \gets$ number of times slots to activate nodes ($k \leq t$)
\STATE $best \gets \infty$
\STATE $AvailableTS \gets \{1, 2, ... t\}$
\STATE $nl \gets$ number of locations of interest
\STATE $V \gets AvailableTS \times \{1, 2, ... nl\}$ (Cartesian product)
\STATE $U \gets \emptyset$
\STATE $best\hat\beta \gets \emptyset$
\STATE $ns \gets$ number of sensors available
\STATE $LA \gets$ sensor locations available
\STATE $LS \gets \emptyset$
\FOR{$i \gets 1$ to $k$}
\STATE $bestJ \gets \emptyset$
\STATE $bestL \gets \emptyset$
\FOR{$j \gets 1$ to $t - i + 1$}
\STATE $ts \gets AvailableTS(j)$
\FOR{$l \gets 1$ to size $LA$}
\STATE $X \gets U \cup \{ts,LA(l)\}$
\STATE $Y \gets V \smallsetminus X$
\STATE $Xdat \gets data(X)$
\STATE $Ydat \gets data(Y)$
\STATE $\hat\beta \gets (Xdat^TXdat)^{-1}Xdat^TYdat $
\STATE $\hat{Ypred} \gets Xdat\hat\beta$
\STATE $err \gets 0$
\FOR {$l \gets 1$ to size $Y$}
\STATE $err \gets err + \frac{1}{n}\sum_{m=1}^n(\hat{Ypred}_{l,m} - Ydat_{l,m})^2$
\ENDFOR
\IF{$err < best$}
\STATE $best \gets err$
\STATE $bestJ \gets j$
\STATE $bestL \gets l$
\STATE $best\hat\beta \gets \hat\beta$
\ENDIF
\ENDFOR
\ENDFOR
\STATE $U \gets U \cup \{AvailableTS(bestJ),LA(bestL)\}$
\STATE $AvailableTS \gets AvailableTS \smallsetminus AvailableTS(bestJ)$
\STATE $LS \gets LS \cup bestL$
\IF{size $LS = ns$}
\STATE $LA \gets LS$
\ENDIF
\ENDFOR
\end{algorithmic}
\end{algorithm}

\subsection{Placement and Scheduling Algorithm}
\label{algo}

Algorithm \ref{cde2} represents the pseudo code for selecting sensor locations along with their associated time slots. In lines 1 to 11, the algorithm is initialised. At this point, all of the time slots will be available. The variables $best$ and $best\hat\beta$ are used to keep track of the minimum sum of MSPEs and the best least squares estimator obtained thus far respectively. The set $U$ is used to represent the selected set of sensor locations and associated time slots and the set $V$ is used to represent the set of locations and time slots where predictions are to be made. $k$ represents the number of sensor activations that are to be made. The sets $LA$ and $LS$ represent the sets of available sensor locations and selected sensor locations respectively. $LS$ differs from $U$ in that the members of the set are not associated with time slots.

In lines 12 to 42, the algorithm enters a loop to select sensors for $k$ time steps. Within this loop, the algorithm keeps track of the best sensor and associated time slot with the indices $bestL$ and $bestJ$ respectively. The algorithm enters an inner loop of all available time slots, or slots that were not previously chosen. For each time slot, each available sensor location, when used in conjunction with the previously chosen sensor locations and associated time slots $U$, is evaluated in turn with regard to the sum of the MSPEs for $Y$, which represents all locations of interest at all time slots barring $U$ and the currently considered sensor location and time slot. If the sum of the MSPEs is less than the best value obtained thus far, in lines 28 to 33, the relevant variables are updated. The sensor location and associated time slot with the lowest sum is greedily chosen (added to the selected set $U$) and the time slot is removed from the set of available time slots (lines 36 and 37). In line 38, the chosen location is added to the set of previously selected locations. If the set's size is equal to the number of sensors available, the set of available locations is changed to be the set of selected locations (lines 39 to 41). This ensures that, subsequently, only locations that were previously chosen to activate sensors can be considered for selection in that there are no additional sensors to activate at new locations. The process continues in this way until $k$ time slots and associated sensor locations have been chosen.

Let $A$ be the set of locations of interest. Let $T$ be the set of time slots $\{1, 2, 3, ... TS\}$, where $TS$ is the window size. Let $V = T \times A$ be the Cartesian product of $T$ with $A$. For each chosen time slot $i$, the node $nd$ chosen is that which, when used with the previously selected set $U$ of sensors for other time slots within the window, minimises the sum of the approximate MSPEs for all locations of interest at all time slots $V$ barring the previously selected sensors for the time slots and the currently considered sensor for the currently considered time slot $V \smallsetminus \{\{i,nd\} \cup U\}$. 

When evaluating the MSPEs, a least squares estimator $\hat\beta$ and sensor readings from the training data $Xdat$ for the currently considered location and the data from the locations from previously selected time slots are used to predict values $\hat{Ypred}$ for the $V \smallsetminus \{\{i,nd\} \cup U\}$ locations and time slots using the equation: $\hat{Ypred} = Xdat\hat\beta$. Subsequently, the MSPE values are determined using the predicted values and the training data. Once the algorithm completes operation, the set of selected locations and associated times will be in the set $U$ and the least squares estimator $best\hat\beta$ associated with $U$ will be used to predict values when data is received from the network.

\subsection{Node Activation}

Algorithm \ref{cde2} operates offline. Once the activation schedule has been determined, it is stored in an array in program memory on the nodes when the nodes are programmed. Alternatively, the schedule could be transmitted to the network while the nodes are active. 

When nodes are in operation, they use the schedule, in conjunction with the sequence number and the count of the number of floods (see Subsection \ref{oon}), to determine when they are to initiate floods. After $t$ time slots have elapsed, the node activation sequence repeats and continues to repeat every $t$ time slots.

\section{Scalable Data Collection}
\label{dir}
In this section, the STAIR many-to-one data collection framework is discussed. The STAIR data collection framework chooses a subset of nodes to activate during flooding to reduce the total energy requirements so  that the performance in terms the amount of data received at the sink is greater than that of alternative approaches, such as Glossy and the LPB \cite{ferrari2012low}, in many-to-one scenarios. Due to the lower total energy requirements, a greater number nodes continue operating and remain connected for longer. With data collection frameworks, such as the LPB, all nodes must activate for all floods. As such, all nodes within the network will have the same workload independently of their location with respect to the sink. Given that the discharge rates of batteries vary, this leads to a situation whereby the longer the network is active, the more nodes (possibly far from the sink) will run out of energy. STAIR alleviates this problem in reducing the total energy requirements and for the time that the nodes close to the sink are active, more of the other nodes will also be active. Thus, the network will produce more data. 

When using constructive interference for flooding, data sent from nodes traverse multiple paths concurrently. This improves the performance because data loss is reduced in that (1) the signal is strengthened due to constructive interference and (2) redundancy is introduced in packet traversal. In order to determine a subset of nodes to activate within a flood with STAIR and to still avail of these benefits, the idea of a width parameter is introduced. The width parameter takes an integer value and has a controlling effect on the number of nodes to activate. The higher width parameter value, the greater the number of nodes will take part in floods and, thus, the more redundancy is introduced in terms of packet traversal and the more constructive interference is produced. If the value of the width parameter is quite low, the performance, in terms of the amount of data received at the sink, will degrade over time due to node failures and topology changes. The drawback of having a very high value for the width parameter, however, is that the network will have a higher overhead in terms of energy consumption.

STAIR goes through 3 phases in determining the sets of nodes to activate in floods. First, data is collected from the network, which is used in the construction of a tree that connects all nodes to the sink. Second, the tree is used to determine the sets of nodes that will participate in floods when the value of the width parameter is equal to 1. Third, the active sets are augmented with additional nodes for larger width parameter values.

\subsection{Tree Construction}
\label{tree}

STAIR makes use of topology information with regard to the link qualities between nodes in the network. In order to obtain this information, initially, profiling is performed whereby nodes individually transmit data to all other nodes in the network and link quality statistics are gathered. Once the link quality statistics have been obtained, a connectivity graph of the network is constructed. Using the connectivity graph, a minimum hop tree, routed at the sink, is constructed. The objective of this part of the algorithm is to construct an array, which represents the parents of all nodes in the tree, and which is used in determining the set of sensors to activate during floods initiated by different nodes.

\subsection{Determining Initial Active Sets}
\label{inita}

For each node in the network, the algorithm determines an initial active set of source nodes that the node must activate for when the source nodes are initiating floods. Using the parent array (see Section \ref{tree}), the algorithm determines the values of the set array for all nodes within the network for the case when the width parameter is equal to 1. The active set is necessary as, with STAIR, nodes within the network must be aware of all of their descendants' identification numbers, rather than their parents' identification numbers. This is because nodes must have information with regard to when they are required to activate so as to relay data from their descendants.

Given a particular time slot and transmission schedule, nodes use the elements of their active set to determine if they must participate in the flood for the source node scheduled to transmit data at that time. That is, when the width parameter is equal to 1, a node's active set elements include the node's identification number along with the identification numbers of the node's descendants in the minimum hop tree.

\subsection{Augmenting Active Sets}

When the width parameter is greater than 1, additional nodes are added to the initial active sets. If the width parameter is 2, the algorithm first constructs the initial active sets as discussed in Section \ref{inita}. Subsequently, it determines an additional node to activate by taking the product of a given source node's link quality to the candidate node and candidate node's link quality to the source node's parent in the shortest hop tree, choosing the candidate that has the highest value for the product. This process is then repeated for the source node's parent, choosing the best node to add to strengthen the link, or introduce redundancy, between the source node's parent and the source node's parent's parent. This process is repeated recursively up the tree and is then repeated starting from the other remaining source nodes. 

When the width parameter is 3, the process begins in the same way as when the width parameter was 2. After the algorithm has recursed up the tree and the best additional node with respect to the link qualities between the node and the sink and the node and the sink's child has been selected, the process starts again from the next source node. The process again selects the best additional nodes, not currently active for the source node's flood, with respect to the link qualities of parents and children in the minimum hop tree. For width parameter values greater than 3, the process continues to operate in a similar manner.

\subsection{Network Operation}
\label{oon}

When the algorithm completes execution, the active set is converted to a bitmap. Subsequently, this bitmap is stored in program memory of the nodes. When the nodes begin operating, they extract the identification numbers of the source nodes they must activate for in floods when the source nodes are acting as the initiators. 

Different source nodes periodically act as the initiator of floods by looping through an activation schedule. The activation schedule is produced either by Algorithm \ref{cde2}, discussed in Section \ref{select}, or using an alternative approach such as round robin scheduling, whereby each node transmits its data in turn. Once every $N$ floods, the sink node transmits a reference packet to all nodes within the network to ensure that all nodes remain time synchronised regardless of whether they have been active in one of the previous $N-1$ floods or not. This packet also serves as a reference point by which all floods not initiated by the sink are synchronized off. That is, floods that were not initiated by the sink use the reference point plus an offset value to determine the times at which nodes should activate to send and receive data. For example, if the reference point was 1 second and the offset value was 250 milliseconds, the first flood would occur at 1.25 seconds, the second flood would occur at 1.5 seconds, and so on for the remaining $N-3$ floods, until the sink flood. The packet also contains a sequence number, which represents the number of floods initiated by the sink. The nodes use the sequence number in conjunction with the node identification numbers obtained from the bitmap and a count of the number of floods that have occurred since the packet was received from the sink to determine the time slots in which to participate in floods.

During a flood, STAIR operates in a similar way to Glossy, but only using a subset of nodes and activating nodes far from the sink less frequently. The nodes iterate through a repetitive sequence of states: {\bf Wait} $\rightarrow$ {\bf Receive} $\rightarrow$ {\bf Transmit}. All nodes, apart from the source node of the flood, begin in the wait state, whereby they turn on their radios and wait to receive data from neighbouring nodes. The source node starts in the transmit state and begins sending data. Once a node begins to receive data, it transitions to the receive state. If at this point, there is a problem with packet reception, for instance if the packet is corrupt, the node transitions back to the wait state. Once the packet has been completely received, the node transitions to the transmit state and begins sending data. Once the packet has been sent, the node transitions back to the wait state provided the number of packets it has transmitted in the flood is less than some value $N$, which represents the maximum number of transmissions a node can make in a flood. Otherwise, the node transitions to an off state. 

When flooding data, the nodes maintain a relay counter that is used in conjunction with a slot length (time between the consecutive start times of packet transmissions in the flood) to determine the time at which the flood was initiated or reference time, which receiver nodes use to synchronise their clocks with the initiator\footnote{With STAIR, nodes are time synchronised with the sink. Prior to a flood initiated by some node other than the sink, the nodes store their reference times with respect to the sink and subsequently restore this value once the flood has completed.}. The following is an example of how flooding operates. Initially, the initiator starts flooding a packet to the network with the counter set to 0. Nodes within communication range of the initiator receive this packet and update the counter to 1. The nodes that received the packet then begin concurrently transmitting data. The information in the packets transmitted by all nodes is the same in that they have all set the counter value to 1. The neighbours of the nodes that received the initial packet, including the initiator, receive the current packet, update the counter value to 2, and begin transmitting. The process continues until all nodes participating in the flood have transmitted the packet $N$ times.

When using STAIR, with suitably chosen values of the width parameter, topology information does not significantly degrade over time (see Section \ref{results}). This is due to the use of constructive interference and the redundancy with regard to packet traversal. As such, after the initial profiling, the nodes can be programmed with the topology information, in the form of the bitmap representation of the active array, and left to operate over extended periods of time without having to perform profiling again. Alternatively, this information could be transmitted to the nodes from the sink on a very low duty cycle. 

\section{Analysis and Discussion}
\label{disc}

Determining the locations and times to activate sensors is a form of subset selection problem, which, in general, is NP-hard. Heuristics are often adopted when addressing NP-hard problems in the creation of algorithms that can operate within reasonable time and memory constraints, but the problem is that there are often no guarantees in terms of the performance of such approaches with regard to the quality of the selections, or search decisions, made when compared to the optimal. As noted in \cite{ICML2011Das}, however, greedy subset selection algorithms that optimise over objective functions that are submodular or approximately submodular, such the optimisation over MSPEs discussed in Section \ref{select}\footnote{In Algorithm \ref{cde2}, the MSPEs are minimised, but this is equivalent to maximising the negative MSPEs, which is approximately submodular.} or the $R^2$ goodness of fit statistic, perform close to optimal and with bounded quality guarantees, even in cases whereby the data is highly correlated.

Submodularity captures the idea of diminishing returns. Consider the selection problem discussed in Section \ref{select}, which exhibits diminishing returns. The larger the currently selected set $B$ of sensor locations and times is, the less the utility will increase, in terms of the coverage, by activating additional sensors at additional times. Formally, $f\colon 2^S \to \mathbb{R}$ is submodular if for $A \subset B \subset S$ and $x \in S$, $f(A \cup \{x\})-f(A) \geq f(B \cup \{x\})-f(B)$. A fundamental result, presented in \cite{nemhauser1978analysis}, proves that in the worst case, the performance of the greedy algorithm when maximising over a submodular function subject to a cardinality constraint will be $1 - 1/e$ of the optimal, which is the best bound that can be obtained for a polynomial time algorithm unless $P=NP$. In general, the algorithm will perform better than the worst case $1 - 1/e$. \cite{ICML2011Das} provides a more general bound of $1 - 1/e^{\gamma}$ for approximately submodular functions where $\gamma$ is a submodular ratio that represents a measure of how close to submodular a set function is. 

One of the primary advantages of STAIR, over architectures that use complete flooding as a communication primitive, is that it keeps more nodes in the network operating prior to the nodes with the highest workload depleting their energy resources and, as such, produces more data. To quantify this, consider the number of messages transmitted and received, or the total energy consumption, as the network size increases. With architectures, such as Glossy, with each additional node, the energy consumption increases by O($n$), using asymptotic notation and where $n$ represents the number of nodes in the network. The total energy consumption will be O($n^2)$ (O($n$) for each node). With STAIR, with each additional node, the consumption increases by O($m$), where $m$ is the average number of hops from each node to the sink, which will be significantly smaller ($m << n$) than O($n$) for large networks. Thus, STAIR, with a total energy consumption of O($nm$) represents a more scalable approach (see Section \ref{nodelt}).

\section{Results}
\label{results}

The following experiments were conducted using the Indriya testbed \cite{doddavenkatappa2012indriya,appavoo2019indriya}, which is deployed over 3 floors of the School of Computing of the National University of Singapore. In the Indriya deployment, 17 nodes had temperature sensors attached. 

\subsection{Sensor Placement and Scheduling Results}

To evaluate Algorithm \ref{cde2}, over 94000 samples from the temperatures sensors in the Indriya deployment were used. In this experiment, the objective was to choose $k$ sensors and associated time slots within a window and to make predictions for all of the other time slots and sensor locations not selected. Current sensor selection algorithms \cite{krause2008near,Clark2016,muldoon2025sensor,shamaiah2010greedy,sakiyama2016} do not consider scheduling. Thus, it is reasonable to assume that in situations whereby data can be only flooded from a single sensor per time slot, and in cases where no schedule is specified, that a round robin or circular approach would be adopted whereby each sensor acts as the initiator node and transmits its data in turn.

\begin{figure}
\begin{center}
\includegraphics[scale = 0.55]{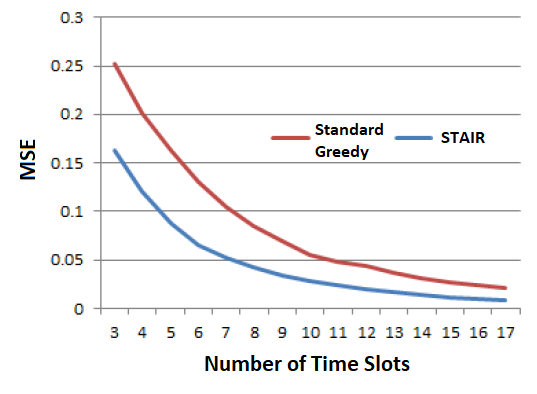}
\end{center}
\caption{Average MSE for a standard greedy approach to sensor selection and STAIR given a fixed number of time slots}
\label{exper6}
\end{figure}

In Figure \ref{exper6}, the STAIR placement and scheduling algorithm (Algorithm \ref{cde2}) is compared to a standard greedy sensor selection algorithm used in \cite{krause2008near,Clark2016,sakiyama2016}. It should be noted that the Algorithm \ref{cde2} is also greedy, in the sense that it does not perform backtracking, but it differs from the standard greedy approach in that it takes into account time and the frequency of sensor activation.  Given a set of available locations $V$ and a set of selected sensor locations $S$, the standard greedy algorithm uses a decision criterion to determine the best location $x \not\in S$ in $V$ until $k$ locations have been chosen. In this experiment, the sum of the MSPEs between $S \cup x$ and the sensor locations associated with $V \smallsetminus {S \cup x}$ is minimised, but an alternative decision criterion could also be used, such as maximising the mutual information of a Gaussian Process \cite{krause2008near,Clark2016} or using a method based on the graph sampling theorem \cite{sakiyama2016}. The algorithm is initiated with $S = \emptyset$ and continues operating until $k$ locations have been selected. With the standard greedy approach, there is no repetition of sensors in that the same sensor cannot be chosen more than once. Given that standard approach does not consider scheduling, the sensors are uniformly timed to flood their data. For example, if there are 10 selected sensors and the window size is 20, the sensors flood their data at times 1, 3, 5, ... 19. With the first selected sensor flooding at time 1, the second at time 3, and so forth for the remaining sensors.

%
%

For results presented in Figure \ref{exper6}, the Mean Squared Error (MSE) of the predictions was calculated for the transmission schedules produced by both the standard greedy selection algorithm and Algorithm \ref{cde2}. A fixed window size of 20 was used and the number of time slots available was varied. The number of time slots available represented the value of $k$ in the standard greedy algorithm. For $n$ predicted values $\hat{y}$ and $n$ true values $y$, the MSE is defined as follows: $\frac{1}{n}\sum_{i=1}^n(\hat{y_i} - y_i)^2$.

Figure \ref{exper6} illustrates that using Algorithm \ref{cde2} leads to an average improvement of 51\%, and up to a 59\% improvement when the number of selected nodes is 16, over the standard greedy approach. As can be seen from Figure \ref{exper6}, the graphs for both the Algorithm \ref{cde2} and the standard greedy selection algorithm are close to convex. When there are few sensors selected, the average MSE decreases by quite a lot with the selection of additional sensors, but when there is a larger number of sensors selected, the average MSE does not decrease by as much. Equivalently, the  problem of selecting the set of sensors that maximise the negative sum of the MSPEs will produce a graph that is close to concave, which is an indicator that the function is exhibiting approximate submodular behaviour. In such cases, greedy algorithms, such as Algorithm \ref{cde2}, perform close to optimal even in situations where there is a high degree of cross correlation \cite{ICML2011Das}. The standard greedy algorithm \cite{krause2008near,Clark2016,sakiyama2016} did not perform as well as Algorithm \ref{cde2} in that it did not take temporal considerations into account or allow for the same locations to be chosen at different time slots. That is, the decision space considered was more restrictive.

\begin{table}[tbp]
\centering
\caption{Number of packets transmitted}
\label{exper2}
\begin{tabular}{|l|l|l|l|l|l|}
\hline
Time & S1 & S4 & S5 & S10 &  Glossy \\
\hline
4 Day Delay  & 65816 & 196178 & 218317 & 248915 &  2260495 \\
Immediate & 61384 & 112712 & 190587 & 236907 & 2248912 \\
\hline            
\end{tabular}
\end{table}

\subsection{Communication Overhead Results}
\label{tenergy}

To evaluate the communication overhead of the STAIR data collection framework, over 34000 floods were performed using 135 nodes in the Indriya testbed. In these experiments, the nodes were wired for power and, thus, did not deplete their energy resources. Section \ref{nodelt} provides the results for the amount of data received for when nodes have a limited lifetime.

The results for the total number of packets that were transmitted within the network are presented in Table \ref{exper2} where S1, S4, ... refer to STAIR with width parameters 1, 4, and so forth. The packet transmission parameter, in these experiments, was equal to 5. That is, in each flood that a node took part in, it transmitted 5 packets, which is the default setting for the reference implementation of Glossy. Floods were initiated once every 250 milliseconds and the reference packet from the sink (see Section \ref{dir}) was transmitted once every 8 floods. The experiments were performed immediately after the profiling phase and again after 4 days without re-performing profiling.

As can be seen from Table \ref{exper2}, there is a substantial reduction in the total communications overhead for STAIR when compared to Glossy. For example, when the width parameter is equal to 5, the total number of packets transmitted will be reduced by over 90\%. The radio activation time to receive messages will also be substantially reduced as, with STAIR, it is proportional to the number of packets transmitted and nodes only activate to relay data in a subset of floods. The communication overhead for STAIR after 4 days does not change significantly, which indicates that the topology information has not become stale.
 
\subsection{Node Lifetime Results}
\label{nodelt}

To quantify the node lifetimes for STAIR, a simulation was performed where data from the Indriya deployment was used to estimate link qualities and determine the network topology. The width parameter for STAIR was taken to be 10. Battery duration times were approximated using the exponential distribution. The exponential distribution takes a parameter $\lambda$, which represents a function of the transmission rate.
\begin{figure}
\begin{center}
\includegraphics[scale = 0.8]{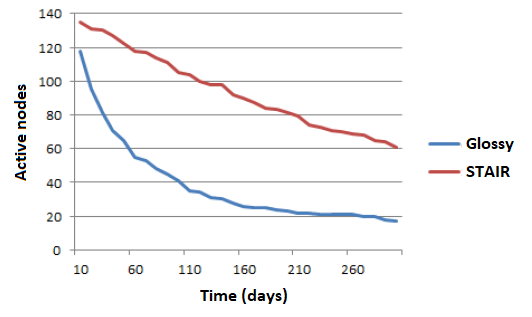}
\end{center}
\caption{Number of active nodes over time}
\label{exper7}
\end{figure}

$\mathrm{E}[X] = \frac{1}{\lambda}$ is its expected value. The expected value represents the average time at which nodes fail. Thus, the larger the transmission rate, the shorter the expected lifetime of the nodes will be on average. In the case where all nodes were active and had the same workload, the expected value was taken to be 40 days. As nodes fail, however, the overall workload  and the transmission rates of the other nodes in the network decreases, which results in the network operating for longer than if transmission rates remained the same. See Figure \ref{exper7} for the number of nodes that remain active over time.

\begin{figure}
\begin{center}
\includegraphics[scale = 0.8]{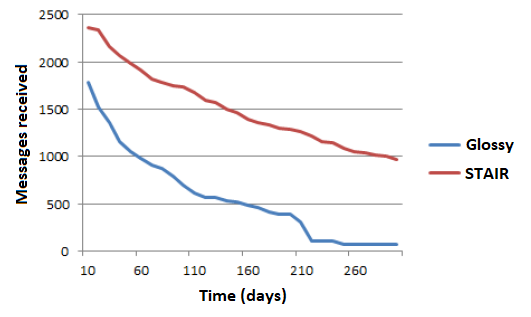}
\end{center}
\caption{Number of messages received over time}
\label{exper8}
\end{figure}

The results of a simulation for the amount of data received at the sink over time are shown in Figure \ref{exper8}. As can be seen from Figure \ref{exper8}, the graph does not follow exactly that of Figure \ref{exper7}. The reason for this is that as nodes fail, parts of the network will sometimes become disconnected and the amount data received will decrease by a greater degree than the reduction in the number of active nodes. As shown by these results, the amount data received by STAIR represents an improvement over 67\% on average.

\section{Conclusion}
\label{concl}
A new class of wireless sensor network has emerged that makes use of constructive interference to enable the development of data collection frameworks that are highly efficient in terms of energy and latency requirements when compared to traditional approaches. This paper presented a sensor placement and scheduling algorithm that has been specifically designed for use with flooding architectures and determines a transmission schedule given flooding as a communication primitive. The algorithm differs from prior work in sensor networks in that it considers the time at which sensors activate and the frequency of sensor activation in determining where and when is best to transmit sensor readings from. Sensors do not operate with a homogeneous sampling rate and more informative sensors are activated more frequently. The sensor placement and scheduling  problem is NP-hard. Heuristics are often adopted to address NP-hard problems, but there are often no guarantees with regard to the performance of such approaches. The sensor placement and scheduling algorithm, discussed in this paper, however, in optimising over the sum of MSPEs for all nodes, which is an approximately submodular objective function, achieves close to optimal performance and has worst-case quality guarantees. In jointly optimising over both sensor activation times and locations, up to a 59\% improvement was observed when the algorithm was tested on the Indriya deployment. 

STAIR delivers significant savings in terms of energy consumption when used for many-to-one data collection, particularly when the network size is large, in that, in contrast to prior research that avails of constructive interference or concurrent transmission, it does not flood the entire network independently of the nodes' locations. In terms of the total communication overhead, flooding only a subset of the network delivers a reduction of up to 90\%. The reduced communications overhead leads to an improvement of over 67\% on average in terms of the amount of data received at the sink as a greater number of nodes remain active and connected for longer. STAIR requires the use of topology information, which enables nodes to determine whether or not to participate in a flood. This information, however, is quite coarse grained and does not degrade significantly over time.

\bibliography{ref2}
\bibliographystyle{abbrv}
\end{document}